\documentclass[conference]{IEEEtran}
\IEEEoverridecommandlockouts
\usepackage{cite}
\usepackage{amsmath,amssymb,amsfonts}
\usepackage{algorithm}
\usepackage[noend]{algpseudocode}
\usepackage{graphicx}
\usepackage{textcomp}
\usepackage{xcolor}
\usepackage{diagbox}
\usepackage{tikz}
\usepackage{float}
\usepackage{array}
\newcolumntype{P}[1]{>{\centering\arraybackslash}p{#1}}
\newcommand*\circled[1]{\tikz[baseline=(char.base)]{
            \node[shape=circle,draw,inner sep=2pt] (char) {#1};}}
\def\BibTeX{{\rm B\kern-.05em{\sc i\kern-.025em b}\kern-.08em
    T\kern-.1667em\lower.7ex\hbox{E}\kern-.125emX}}
\newcommand{\spaarc}{SPAARC}
\begin{document}

    \title{SPAARC: Spatial Proximity and Association based prefetching for Augmented Reality in edge Cache}

    \author{\IEEEauthorblockN{Nikhil Sreekumar, Abhishek Chandra, Jon Weissman}
    \IEEEauthorblockA{\textit{University of Minnesota, Twin Cities} \\
    MN, USA \\
    \{sreek012, chandra, weiss039\}@umn.edu}}

    \maketitle

    \begin{abstract}
        Mobile Augmented Reality (MAR) applications face performance challenges due to their high computational demands and need for low-latency responses.  Traditional approaches like on-device storage or reactive data fetching from the cloud often result in limited AR experiences or unacceptable lag. Edge caching, which caches AR objects closer to the user, provides a promising solution. However, existing edge caching approaches do not consider AR-specific features such as AR object sizes, user interactions, and physical location. This paper investigates how to further optimize edge caching by employing AR-aware prefetching techniques. We present \spaarc, a Spatial Awareness and Association-based Prefetching policy specifically designed for MAR Caches.  \spaarc{} intelligently prioritizes the caching of virtual objects based on their association with other similar objects and the user's proximity to them. It also considers the recency of associations and uses a lazy fetching strategy to efficiently manage edge resources and maximize Quality of Experience (QoE).

        Through extensive evaluation using both synthetic and real-world workloads, we demonstrate that \spaarc{} significantly improves cache hit rates compared to standard caching algorithms, achieving gains ranging from 3\% to 40\% while reducing the need for on-demand data retrieval from the cloud. Further, we present an adaptive tuning algorithm that automatically tunes \spaarc{} parameters to achieve optimal performance. Our findings demonstrate the potential of \spaarc{} to substantially enhance the user experience in MAR applications by ensuring the timely availability of virtual objects.
    \end{abstract}

    \vspace{-3mm}
    \section{Introduction}
\label{sec:intro}
    Mobile Augmented Reality (MAR) \cite{bib:arsurvey1} is a transformative technology with applications across diverse domains, from gaming \cite{bib:pgorest} and education \cite{bib:eduar} to healthcare \cite{bib:healthar} and manufacturing \cite{bib:manuf}. However, its reliance on data-intensive computations and need for low latency presents unique challenges.
    
\begin{figure}[t]
        \centering
        \includegraphics[scale=0.1]{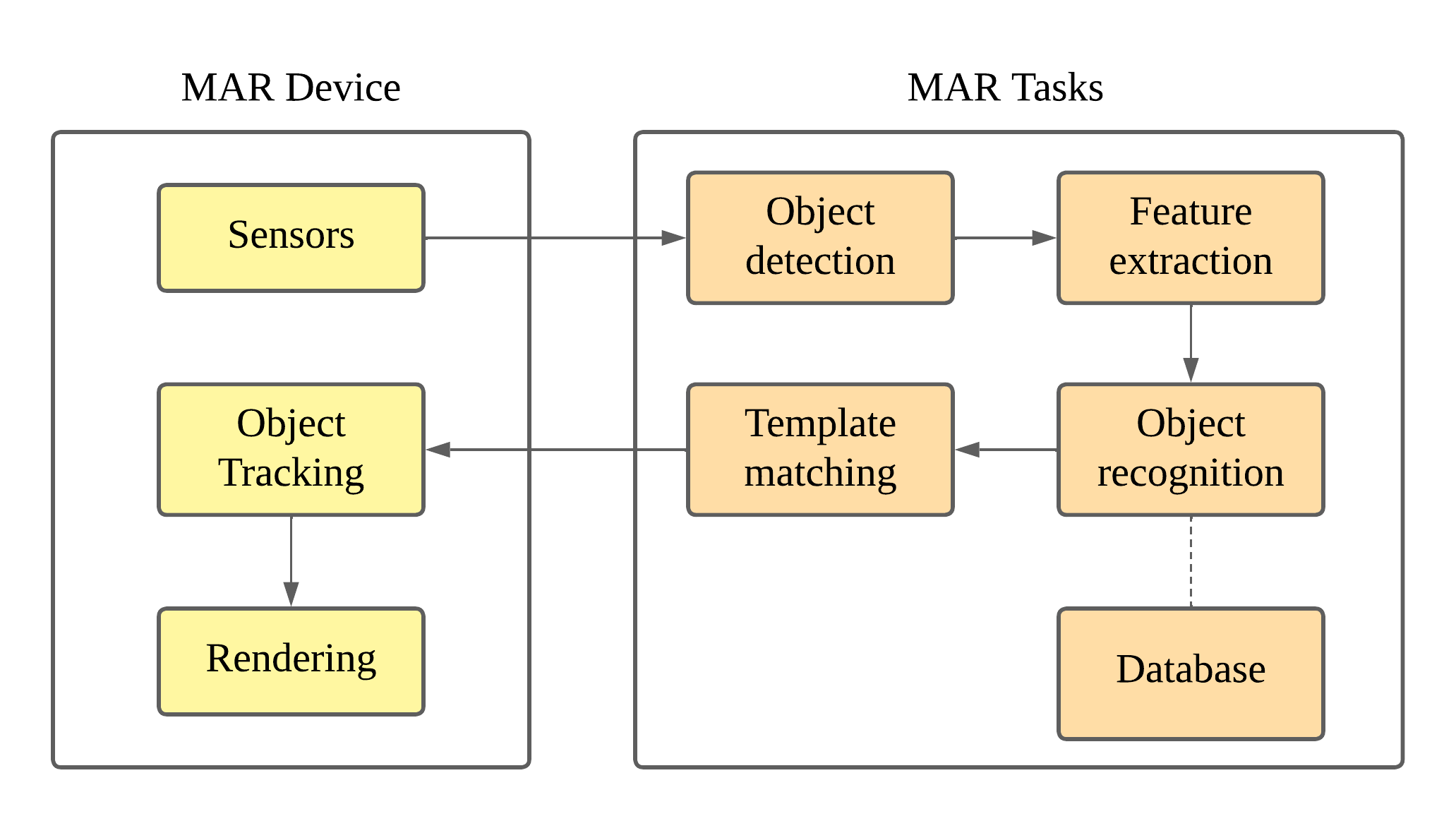}
        \caption{A Typical Mobile Augmented Reality (MAR) Pipeline}
        \label{fig:marpipeline}
        \vspace{-7mm}
    \end{figure}

    Figure \ref{fig:marpipeline} illustrates a typical MAR pipeline, comprising MAR Device (end-user device) and MAR Tasks (compute-intensive tasks and their associated resources) \cite{bib:marpipeline}. These tasks can be distributed across the user device, edge, and/or cloud based on resource availability. The key stages in this pipeline include (1) Object detection and feature extraction: identifying potential target objects in a camera frame and extracting their unique features; (2) Object recognition and pose estimation: matching extracted features to a database to identify the object and estimate its real-world pose, potentially aided by template matching \cite{bib:tempmatch}; (3) Object tracking: continuously monitoring the identified object across frames to optimize performance \cite{bib:objtrack}; and (4) Image rendering and virtual content overlay: retrieving and overlaying corresponding virtual content onto the real-world object, achieving the core AR functionality. Efficient data management, particularly the storage of virtual object content, is crucial for optimal performance in this pipeline.

    Traditional approaches to managing virtual object data in MAR, such as on-device storage \cite{bib:ondevice} or cloud offloading \cite{bib:overlay}, face limitations.  Increasingly complex and interactive virtual objects demand significant storage, rendering on-device storage impractical for lightweight mobile devices.  On the other hand, on-demand fetching from the cloud suffers from high latency \cite{bib:clt1, bib:clt2}, hindering smooth real-time AR experience.
    

    Edge computing offers an attractive alternative for MAR applications \cite{bib:arsurvey1,bib:arvrcache}. Processing data closer to the user at the network edge reduces latency, enabling real-time interaction and a more immersive experience.  Further, edge computing decreases bandwidth demands by reducing data transmission to the cloud, leading to increased efficiency and cost savings.  Offloading processing to the edge also enhances reliability, allowing MAR applications to function even with intermittent network connectivity. Finally, the ability of edge infrastructure to store virtual objects locally further improves application performance and responsiveness.

    Edge caching is a critical technique for latency-sensitive applications, including AR \cite{bib:cachecompar, bib:cachemec,bib:onlinecoll, bib:arvrcache, bib:artactile}.  To achieve immersive AR experience, caching is necessary across various stages of the MAR processing pipeline.  Some existing approaches, such as CARS \cite{bib:carsar}, Precog \cite{bib:precog} and SEAR \cite{bib:sear}, focus on caching image recognition results at the edge, while others advocate for caching virtual objects directly on user devices \cite{bib:localcache}.  Prefetching AR data to further enhance cache hit rates remains an active research area.  Current prefetching techniques primarily address location-based services (LBS) and video content.  In LBS, prefetching typically involves caching objects near the user's location on either the device or an edge server \cite{bib:dreamstore, bib:locar,bib:advlocar}.  For video content, prefetching leverages user interest to cache video segments on nearby edge servers, often in conjunction with techniques like tile caching \cite{bib:explorevr, bib:cubist, bib:coopec, bib:tile}. Edge caching resources, although more abundant than mobile device storage, are still limited and heterogeneous. Therefore, it is challenging to ensure the right set of virtual objects are readily available for augmentation at the edge.
    

    This paper investigates optimizing edge prefetching in AR applications by analyzing user-object interaction patterns and AR-specific properties. Given the context-dependent nature of AR interactions, we argue that effective prefetching strategies should prioritize objects likely to enter the user's field of view (FoV).  We hypothesize that predictable patterns in user access behavior can be leveraged to anticipate future interactions and prefetch relevant virtual objects at the edge, thereby improving rendering performance and enhancing the overall AR experience. 
    
    Based on these insights, we present an {\em MAR edge cache prefetching framework called \spaarc{}} that exploits two key properties for AR object prefetching: (i) {\em object associations} (how likely multiple objects are to be accessed together by a user), and (ii) {\em spatial proximity} (how close an object is to a user's FoV). Our approach uses association mining to determine object associations, and distance thresholding to identify spatial proximity. It further incorporates factors such as recency of associations and lazy prefetching to ensure efficient utilization of edge resources and network bandwidth. We also present an adaptive tuning algorithm to automatically tune the parameters of these algorithms. 
    We note that \spaarc{} is designed to be complementary to the underlying caching policies, and its prefetching can be used on top of any caching algorithm to further improve the cache performance.

    This paper makes the following research contributions:
    
    \noindent$\bullet$ Analysis of 
    the rationale for using object association and proximity to achieve efficient MAR edge caching.
    
    \noindent$\bullet$ The design of \spaarc{}: a cache prefetching framework that utilizes virtual object associations and user proximity to improve edge cache performance, along with an adaptive parameter tuning algorithm.
    
    \noindent$\bullet$ A comprehensive evaluation of \spaarc{}, using both synthetic and real-world traces, showing that \spaarc{} significantly improves hit rates (by 3-40\%) compared to various baseline caching algorithms.

    \section{Motivation and Background}
\label{sec:motive}
    Achieving an immersive mobile augmented reality (MAR) experience presents unique challenges \cite{bib:networkar,bib:locar}. Conventional data caching and prefetching policies, designed for general-purpose applications, might not be optimal for MAR due to its inherent characteristics\cite{bib:archaract}. Compared to other edge-native applications \cite{bib:edgenative}, MAR applications require significant computational resources and sophisticated data management strategies to handle the large volume of spatial data and real-time interactions. 

    \subsection{Motivating AR Scenarios}
    \begin{figure}
        \centering
        \includegraphics[scale=0.13]{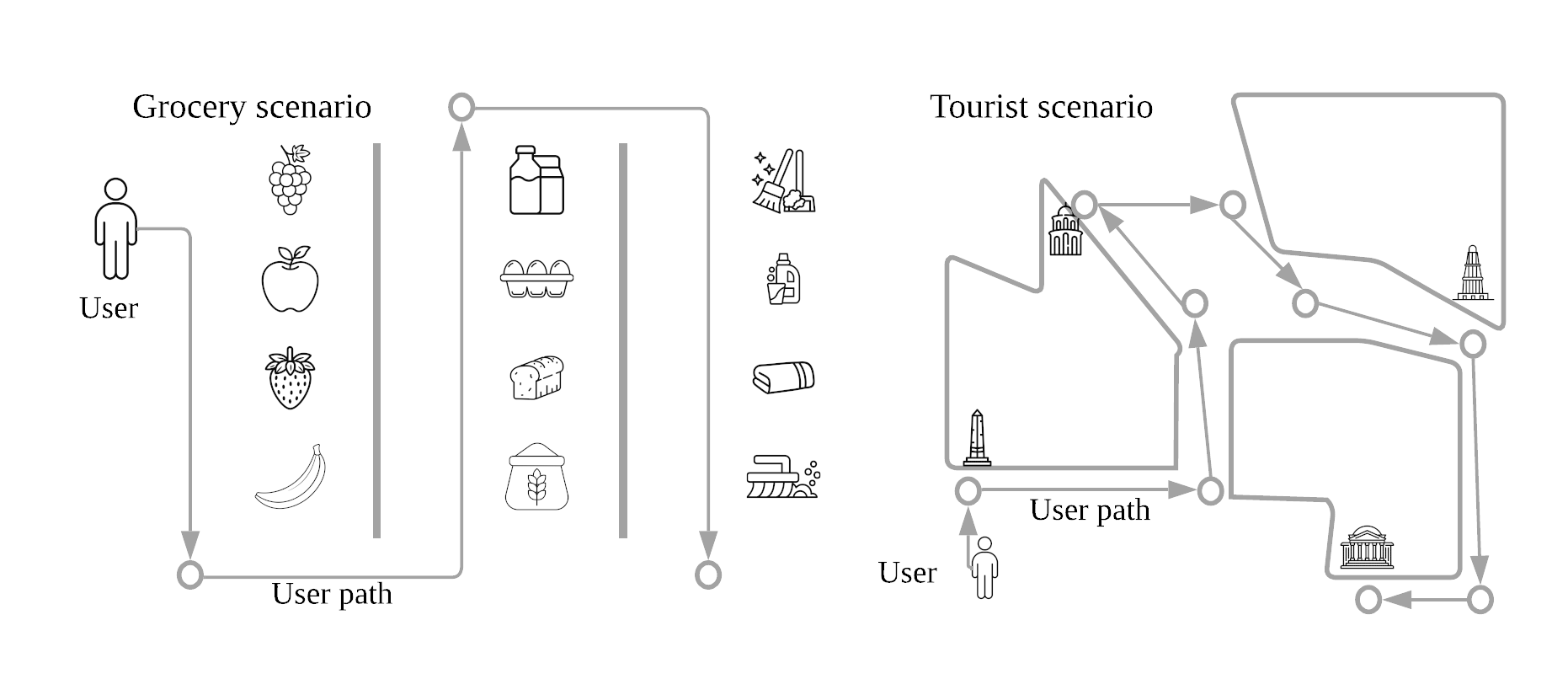}
        \caption{Grocery and tourist scenarios}
        \label{fig:motiveapp}
        \vspace{-6mm}
    \end{figure}
    Figure \ref{fig:motiveapp} shows two example application scenarios for AR. In a grocery store scenario\cite{bib:phara}, users interact with virtual objects of the store items to receive information about their ingredients, manufacturing details, price variations, recipes, and so on. The items are typically placed on racks in aisles based on their categories, and there is a well defined order of placement. The Field of View (FoV) of a user will have multiple items at their location in the store, from which the user might be interested in a few. 
    
    In the case of a tourist scenario, \cite{bib:dublin}, attractions are typically scattered across a region (such as a city block, campus, or neighborhood). The virtual objects may contain historical facts, 3D representations of buildings, informative audio/video, and so on. A tourist would visit the attractions based on multiple factors like user interests, transportation availability, accessibility, guide recommendations, and so on. The FoV of the user in this case would likely consist of only a few items in their vinicity, and the user will have to explore the region more to find the next attraction. 

    Such MAR scenarios face the following challenges:
    
    \noindent$\bullet$ {\em Limited device storage and large object sizes}:  While current virtual objects are typically under 20MB \cite{bib:carsar, bib:objaverse}, increasing interaction complexity and 3D objects will lead to larger object sizes.  Storing all objects locally on the device  becomes impractical as object size and complexity grow.
    
    \noindent$\bullet$ {\em High cloud latency and strict AR latency requirements}: Average cloud latency (around 60ms \cite{bib:clatency, bib:clt1, bib:clt2}) exceeds the sub-20ms threshold required for immersive AR experience \cite{bib:20ms}.  On-demand fetching of objects from the cloud introduces significant latency \cite{bib:delayhits}, negatively impacting user experience.  

\subsection{Edge caching}
    Edge infrastructure is well placed to meet these challenges by expanding on the limited on-device storage capacity, while reducing the user latency. The use of the edge in an MAR context necessitates novel data management policies that consider both AR-specific data characteristics and edge/cloud parameters. 
    
    Recent research on MAR architecture and applications \cite{bib:edgearch, bib:carsar, bib:sear,bib:localcache,bib:slamshare} partitions data and tasks across mobile devices and edge servers to leverage resources at both locations and facilitate data sharing. Caching is crucial for improving AR application performance, and partitioning the cache across devices and edge servers has shown potential for increasing hit rates and reducing latency~\cite{bib:carsar, bib:sear}. 
    
    However, existing approaches often neglect critical AR-specific factors: (1) \textit{Increasing Object Sizes}: Caching strategies must adapt to the growing size and complexity of virtual objects. (2) \textit{Occlusion and User Proximity}: Virtual objects can be occluded by physical objects in AR environments. Additionally, users are more likely to interact with nearby objects. These spatial considerations, along with object access patterns, can inform more effective caching strategies. (3) \textit{Object Relevance and User Associations}: User interactions with virtual objects often exhibit predictable patterns (e.g., a user interacting with a milk carton may subsequently interact with nearby eggs or bread). However, caching strategies must assess the relevance of potential interactions to avoid unnecessary cache pollution.

    \subsubsection*{The need for Prefetch}
        Edge caching offers a promising solution by storing frequently accessed objects closer to the user, reducing cloud traffic and improving access times. However, efficient cache management is crucial given the limited and heterogeneous nature of edge resources.  Traditional cache eviction algorithms (e.g., LRU, LFU, FIFO, etc.) can be employed on the edge cache, but relying solely on reactive retrieval from the cloud upon cache misses can still introduce significant latency. Predictive prefetching based on access patterns can anticipate future requests and proactively cache necessary objects at the edge, reducing cache misses and enhancing user experience at the potential expense of increased network usage.
        
    \vspace{-1.5mm}
    \subsection{Associations and Spatial Proximity}
        Association rule mining (ARM) \cite{bib:arm}, a technique widely used in recommender systems to predict user-item interactions, offers a promising approach for AR prefetching. ARM has proven successful in various domains, including e-commerce \cite{bib:phara}, tourism \cite{bib:dublin}, fraud detection \cite{bib:fraud}, and social network analysis\cite{bib:sna}. For example, in online grocery shopping, purchasing milk and eggs often triggers recommendations for bread and jam. However, the direct application of ARM for prefetching in AR remains an underexplored area.  The rule generation process, particularly with frequent updates, can be computationally expensive, posing a challenge for latency-sensitive AR applications. This work explores adapting ARM to the specific needs of AR, aiming to achieve efficient prefetching while minimizing computational overhead. While ARM effectively identifies related virtual objects, it may not prioritize those most relevant to the user's current location. Prefetching objects far from the user's field of view (FoV) provides limited value in AR. Therefore, this work emphasizes incorporating spatial awareness into the prefetching process. By prioritizing objects near the user's FoV, we aim to optimize cache utilization and enhance the user experience.

    \section{\spaarc}
\label{sec:spaar}
We propose \spaarc, a prefetching framework that addresses the limitations of existing edge caching approaches for MAR applications. It incorporates both the proximity of virtual objects to users and object associations to make informed decisions on edge server caching. Its prefetching policy is complementary to the underlying caching algorithm employed by the edge cache, and is meant to enhance the cache performance.

    \subsection{System Architecture}
\begin{figure}[t]
            \centering
            \includegraphics[scale=0.1]{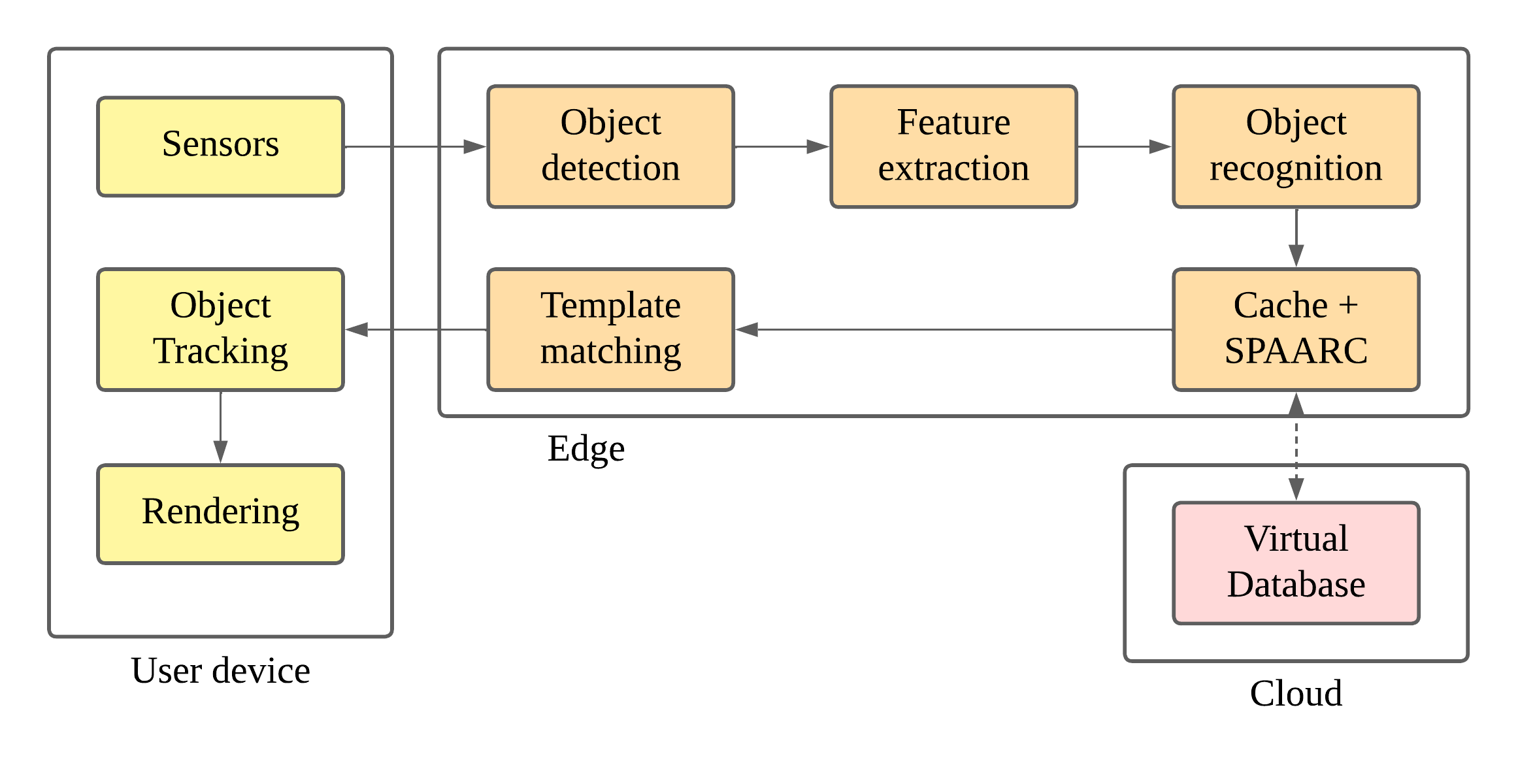}
            \caption{MAR Pipeline with \spaarc{} on the edge cache}
            \label{fig:marspaarc}
            \vspace{-5mm}
        \end{figure}

        Figure \ref{fig:marspaarc} illustrates the MAR pipeline incorporating \spaarc{} into the edge cache.  Sensor data from the user's device is transmitted to the nearest edge server, where video frames are pre-processed and object detection is performed.  Extracted features are then used for object recognition. If a corresponding virtual object exists, the system attempts to retrieve it from the edge cache.  Upon a cache miss, the \spaarc{} algorithm is invoked to identify all necessary objects for prefetching and request them from the cloud, along with the object that caused the cache miss. After receiving the virtual objects, template matching is performed to estimate their pose. This pose information, along with the virtual objects, is sent to the object tracking module on the user's device.  The tracking module identifies objects across frames, and this information, combined with data from the edge server, is used by the rendering module to overlay virtual content onto the user's display.
        

        We next discuss the two main \spaarc{} techniques that drive its prefetching:  association and proximity.
        
        \subsection{Association}
        \label{ssec:spaarc_assoc}
        Intuitively, if a user accesses a virtual object in an MAR scenario, they are likely to soon access related objects (e.g., milk and eggs in the grocery scenario), which should be prefetched to the edge cache for faster access. \spaarc's Association technique generates association rules using {\em Association rule mining (ARM)} \cite{bib:armorig, bib:arm} to predict frequently co-accessed virtual objects based on user interaction history. 
        
        ARM identifies relationships between items in a transaction database by discovering {\em frequent itemsets}: groups of items that frequently co-occur in transactions. The {\em support} $(S)$ of an itemset indicates the proportion of transactions containing that itemset, with frequent itemsets exceeding a predefined minimum support threshold. Based on these frequent itemsets, ARM generates {\em association rules} of the form  $A \implies B$, where $A$ (antecedent) and $B$ (consequent) are disjoint itemsets. The {\em confidence} $(C)$ of a rule represents the conditional probability of observing the consequent in transactions containing the antecedent, controlled by a minimum confidence threshold.  The {\em lift} $(L)$ of a rule measures the strength of the association between the antecedent and consequent, quantifying how much more likely they are to co-occur than if they were statistically independent. Support of an itemset $X$ ($S(X)$), confidence of a rule $A \implies B$ ($C(A \implies B)$) and lift of the rule ($L(A \implies B)$) are formally represented as follows:
            \begin{align}
                S(X) & = \frac{T(X)}{T} \\
                C(A \implies B) & = \frac{S(A \implies B)}{S(A)}\\
                L(A \implies B) &= \frac{S(A \cup B)}{S(A) * S(B)}
            \end{align}
            where $T(X)$ and $T$ are the number of transactions containing $X$ and the total number of transactions respectively. We use ARM algorithms to either generate rules a priori from history user interactions or dynamically during the process.

            \begin{figure}
                \centering
                \includegraphics[scale=0.11]{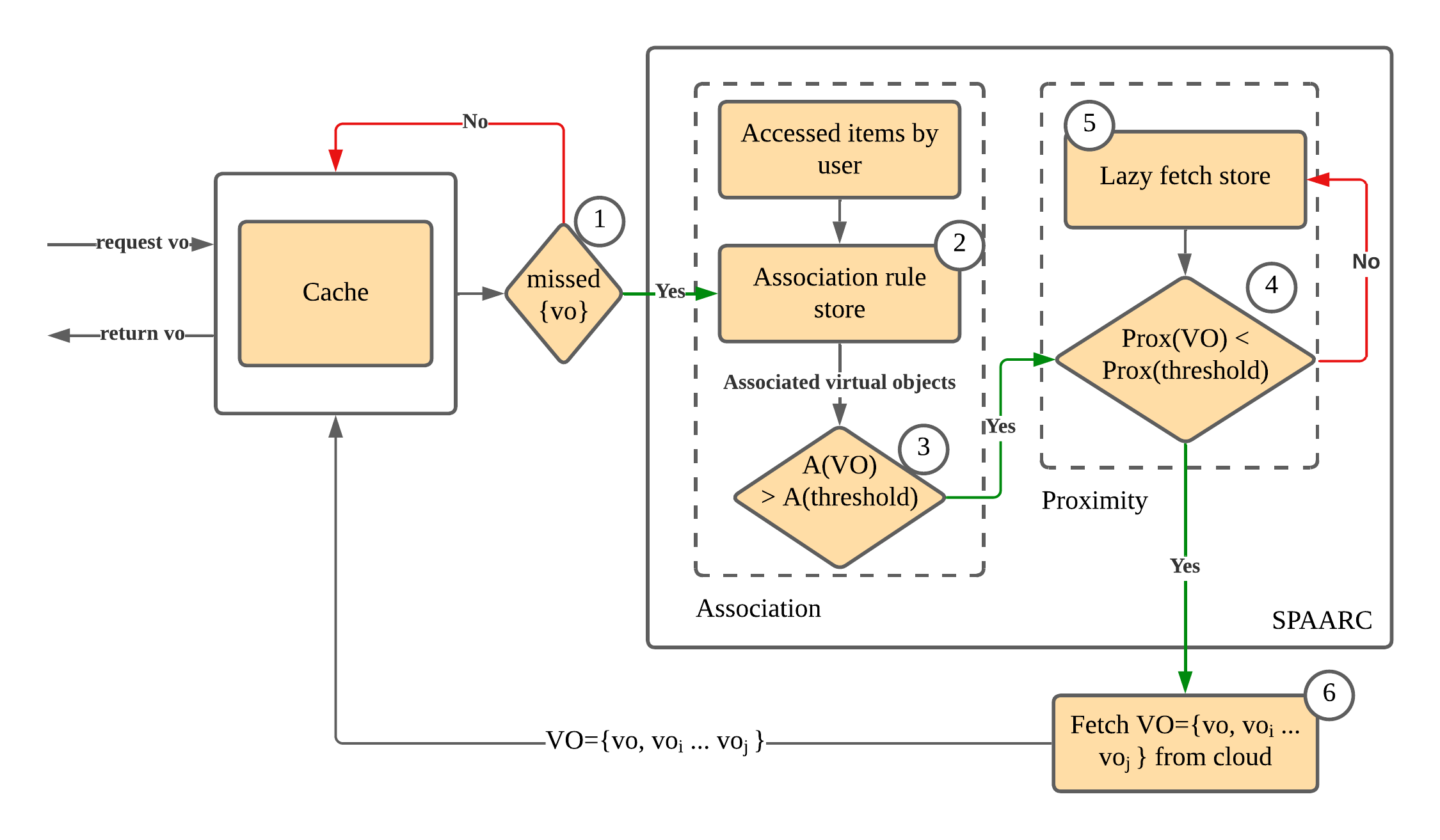}
                \caption{\spaarc{} workflow}
                \label{fig:spaarcworkflow}
            \end{figure}

\spaarc{} uses an ARM algorithm to generate a set of associated objects for prefetching whenever there is a cache miss.
However, 
all such associated objects are not equally relevant for prefetching. Objects that have been accessed more frequently and more recently are more likely to be more useful for caching. Thus, \spaarc{} prioritizes relevant objects within an association set using an {\em association factor} value for each object. This factor reflects the combined influence of frequency and recency of association with the missing object. Objects with higher association factors are prioritized for caching due to their increased likelihood of subsequent user interaction. 

We calculate the association factor ($A$) of a virtual object $vo$ as follows:
            \begin{align}
                A(vo)_{new} & = (F(vo) * \alpha) + A(vo)_{old} * (1 - \alpha)
            \end{align}
            where $\alpha = \frac{2}{1 + window}$, $window$ is the window frame of previous object interactions which are relevant, $F(vo)$ is the number of times the virtual object $vo$ is reference in the window frame. The value of $window$ could be tuned according to the recency requirement. A larger window implies longer trend and a smaller one for shorter trend. If the access patterns are not changing frequently, a larger window size would suffice. Association factor threshold value should be set such that low association objects are filtered. The higher the threshold value, higher the chance that the selected objects are accessed in the near future.

        \subsection{Proximity}
        \label{sssec:prox}
            Since relevant AR objects are based on the field of view of the user, the system only needs to prefetch objects that are in the user's physical vicinity. \spaarc{} incorporates spatial proximity to refine the prefetching list generated by the Association component. Proximity refers to the distance of virtual objects $(Prox(vo))$ from the user's location. For instance, in the grocery store scenario (Figure \ref{fig:motiveapp}), an association rule might suggest prefetching ``milk" alongside ``apples" and ``bananas". However, if the user is not yet in the aisle containing milk, immediate prefetching of the ``milk" virtual object is unnecessary. \spaarc{} prioritizes contextual relevance by employing a {\em lazy prefetching} strategy, deferring the prefetching of ``milk" until the user is in closer proximity. This ensures that prefetched objects are highly relevant to the user's immediate surroundings. A {\em proximity threshold} $(Prox(threshold))$ is used to identify objects within a specific distance from the user. This threshold is domain-specific and depends on the application use case and physical environment. 


             \subsection{SAAP Workflow}
            Figure \ref{fig:spaarcworkflow} illustrates the workflow of the \spaarc{} framework. Upon encountering a cache miss for virtual object $vo$ \circled{1}, \spaarc's Association module utilizes the ARM-generated rules to identify potentially associated objects based on the user's access history and the missing object \circled{2}. To refine this selection and prioritize relevant objects, it uses the association factor for each object. Objects with association factors higher than the Association factor threshold are prioritized for caching due to their increased likelihood of subsequent user interaction \circled{3}.
            Next, the Proximity module filters the selected objects further to those within the Proximity threshold of the user \circled{4}. The filtered out relevant objects are set aside for lazy fetch later when the user moves closer to them \circled{5}.
            Once all the virtual objects to be fetched are identified, the request is send to the cloud \circled{6}. On retrieval, the objects are stored in cache and the missed virtual object is returned. Note that \spaarc{} works in a complementary manner to the cache that continues to use its caching algorithms, e.g., for evicting any objects needed to make space in the cache.

    \subsection{Adaptive Tuning}
        In ARM, selecting the minimum support and minimum confidence thresholds is often domain-dependent, requiring expert knowledge. To automate \spaarc, we focus on tuning the minimum support parameter. Minimum support is prioritized as it directly influences the generation of frequent itemsets, which are fundamental to the subsequent steps in the process. Other parameters can be tuned in a similar manner.

        \begin{table}
                \caption{Algorithm \ref{alg:minsuptune} Notations}
                \label{tab:annot}
                \begin{center}
                    \begin{tabular}{|P{1cm}|p{7cm}|}\hline
                        Notation & Remark\\
                        \hline\hline
                        $\delta$  & hit rate degradation threshold\\
                        \hline
                        $\beta$  &  minimum support\\
                        \hline
                        $\gamma$  & minimum confidence\\
                        \hline
                        $T$ & transactions\\
                        \hline
                        $\zeta$ & lift\\
                        \hline
                        $D$ & number of evenly spaced minimum support values to be selected in the given range that determines the granularity of the search for the optimal minimum support value\\
                        \hline
                        $\kappa$ & kurtosis, statistical measure of the "tailedness" of the distribution of the lift of generated rules\\
                        \hline
                        $\eta$ & threshold for the ratio of the number of association rules to the number of frequent itemsets that helps to control the number of the generated rules\\
                        \hline
                        $\theta$ & used to determine when the distribution of lift has changed significantly, indicating the generation of large number of rules, many of which may be irrelevant\\
                        \hline
                    \end{tabular}
                \end{center}
            \end{table}
            
        \noindent \textbf{Minimum support}: 
            Algorithm \ref{alg:minsuptune} outlines the procedure for tuning the minimum support parameter (Notations described in Table \ref{tab:annot}). The key insight is that the quality of the parameter value (and hence the corresponding association rules) is measured by its impact on the cache performance (hit rate).
            
            The TuneMinSup function dynamically adjusts the minimum support based on cache hit rate degradation. It determines whether to generate new association rule sets, select from existing ones, or continue with the current/next set, depending on the degree of hit rate degradation relative to a predefined threshold.

            The GenARules function identifies the bounds for minimum support and generates $N$ new association rule sets, ranging from low to high support values, based on the last $n$ transactions. The value of $n$ can be adjusted to meet the application's association recency requirements.
        
            The SetARules function selects the appropriate rule set to use. Whenever new $N$ rulesets are generated in the increasing order of minimum support, the function selects the middle ruleset. Otherwise, it goes in the increasing or decreasing minimum support direction, depending on the hit rate degradation. This dynamic adjustment allows the system to adapt to changing access patterns and maintain optimal performance.
        
        \begin{algorithm}
        \footnotesize{
            \caption{Minimum support tuning and ruleset generation}
            \label{alg:minsuptune}
            \begin{algorithmic}[1]
                \Procedure{TuneMinSup}{$\delta, \gamma, T$}
                    \State $hrd \gets getDegradation()$
                    \If {$hrd > 2 * \delta$}
                        \State $GenARules(T, \gamma)$
                    \EndIf
                    \State $SetARules()$
                \EndProcedure

                \Procedure{GenARules}{$T, \gamma$}
                    \State $\{\beta_{low}, \beta_{high}\} \gets getMinSupBound(T)$
                    \State $\{\beta_{temp}\} \gets divideMinSupBound(\{\beta_{low}, \beta_{high}\}, D)$
                    \State $\kappa_{prev}, \kappa_{curr} \gets NULL$
                    \State $\{\beta_{newlow}, \beta_{newhigh}\} \gets \{\infty, -\infty\}$
                    \For{$\beta_{t} \in \{\beta_{temp}\}$} \Comment Largest to smallest minsup
                        \State $fqitemsets \gets genFreqItemsets(T, \beta_{t})$
                        \If{$fqitemsets.size > 0$}
                            \State $arules \gets genARules(fqitemsets, \gamma)$
                            \State $arules \gets arules[\zeta \ge 1]$
                            \If $\frac{arules.size}{fqitemsets.size} > \eta$
                                \State $break$
                            \Else
                                \State $\kappa_{curr} \gets kurtosis(arules[\zeta])$
                                \If {$abs(\kappa_{prev} - \kappa_{curr}) > \theta$}
                                    \State $break$
                                \Else
                                    \State $\beta_{newlow} \gets min(\beta_{newlow}, \beta_{t})$
                                    \State $\beta_{newhigh} \gets max(\beta_{newhigh}, \beta_{t})$
                                    \State $\kappa_{prev} \gets \kappa_{curr}$
                                \EndIf
                            \EndIf
                        \EndIf
                    \EndFor

                    \State $ruleset \gets genARuleSets(T, \{\beta_{newlow}, \beta_{newhigh}\}, $ $\gamma, N)$ \Comment Generate N rulesets and store
                    \State $SetARules()$
                \EndProcedure
            \end{algorithmic}
            }
        \end{algorithm}
        
        In the GenARules function, the initial lower and upper bounds for minimum support are first determined. Since popular items typically represent a small percentage of the total unique items in most datasets \cite{bib:spmf}, the lower bound is set to the average support of all items \cite{bib:minsup}.  From this initial range, $D$ evenly spaced minimum support values are selected. For each value, association rules are generated with a fixed minimum confidence. During rule generation, the process terminates if the rule count ratio exceeds a predefined threshold $(\eta)$, indicating the potential inclusion of irrelevant rules. Otherwise, the kurtosis value is recorded. If the difference in kurtosis between the current and previous rule sets exceeds another threshold $(\theta)$, the process terminates, signaling the generation of a large number of rules with relatively low lift values.  This step helps to prevent overfitting and ensures the selection of meaningful rules. If neither termination condition is met, the current minimum support range is updated. Finally, $N$ rule sets are generated from the identified minimum support bounds and utilized accordingly. This adaptive approach allows for efficient and effective rule generation tailored to the specific characteristics of the workload.
\section{Evaluation}
\label{sec:eval}
\subsection{Experimental setup and methodology}
    We use a combination of simulations and real testbed experiments.
    Our experimental testbed utilizes a 64GB Intel Xeon E5-2620 with 24 cores (end-users), a 16GB AWS Local Zone EC2 t3.xlarge (edge server), and a 16GB AWS EC2 t2.xlarge (cloud server).  Evaluation is performed using four synthetic datasets and a real-world dataset \cite{bib:spmf}. 
    

    We designed an AR workload simulator that replicates key aspects of user behavior and object placement: (1) Environment: Constructs a geographical region with obstacles using collider data \cite{bib:collider}, mimicking real-world scenarios with varying object placement densities . The distance between objects is within the range of [10, 15] unit distance. (2) User Behavior: Models user movement, gaze direction, and interaction time with virtual objects. (3) User Traffic: Simulates user arrivals using a Poisson process and interaction time with each object using a normal distribution, capturing both regular and peak usage scenarios.
    \begin{table}
            \caption{Datasets}
            \label{tab:datasets}
            \begin{center}
                \begin{tabular}{|l||*{5}{p{1cm}|}}\hline
                
                    Dataset & Type & Random Itemset Support & Number of items & Number of users & Scenario\\
                    \hline\hline
                    DS30  & Synthetic & 30\% & 20-100 & 100 - 1k & Library \\
                    \hline
                    DS45  & Synthetic & 45\% & 20-100 & 100 - 1k & Shopping mall\\
                    \hline
                    DS60  & Synthetic & 60\% & 20-100 & 100 - 1k & Zoo/
                    aquarium\\
                    \hline
                    DS75 & Synthetic & 75\% & 20-100 & 100 - 1k & Gallery guided tour\\
                    \hline
                    Grocery & Real & Varying & 169 & 9k & Grocery store\\
                    \hline
                \end{tabular}
            \end{center}
        \end{table}

    \begin{figure*}
            \centering
            \includegraphics[scale=0.5]{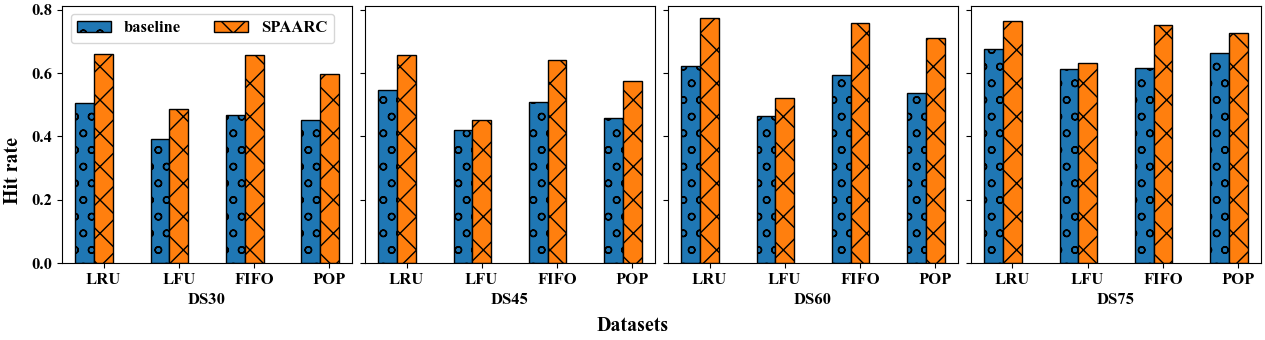}
            \caption{Hit rates across datasets. The top hit rates achieved by \spaarc{} compared to baselines.}
            \label{fig:besthr}
        \end{figure*}
    \subsection{Workloads}
        We utilize both synthetic and real world datasets (Table \ref{tab:datasets}) for our experiments.

        \subsubsection{Synthetic Workloads}
            Existing AR workload datasets often focus on a single user's perspective, limiting their applicability for evaluating caching strategies that consider interactions from multiple users. To address this gap, we leverage the AR Workload Simulator to generate a suite of multi-user workloads. To generate our workloads, we first defined the percentage of unique items that would form these frequent itemsets. Subsequently, we created transactions where the support for these pre-defined itemsets was fixed. It is important to distinguish this ``support" from the minimum support parameter used in frequent itemset generation algorithms. Here, support is solely used for dataset creation, and the resulting dataset might contain additional frequent itemsets beyond the pre-defined ones. We utilize four synthetic datasets (DS30, DS45, DS60, DS75) to evaluate the effectiveness of prefetching under diverse user interaction patterns. The support level for randomly chosen itemsets is systematically varied across the datasets (Table \ref{tab:datasets}). By adjusting the support level, we can assess caching performance under different user access patterns reflected by the frequency of item co-occurrence. Each dataset corresponds to a different real-world MAR scenario, as shown in the table. Depending on the type of experiment we vary the number of users, and the number of items for each dataset. Unless otherwise specified, the default configuration utilizes 100 users, 50 objects and a cache size of 20\% of total size of 50 objects. The size of virtual objects are in the range [10, 15] MB \cite{bib:carsar, bib:objaverse}.

        \subsubsection{Real Workload}
            SPMF \cite{bib:spmf} consists of multiple real world workloads with varying transaction size and item counts. We mainly focus on the grocery dataset for our experiments. It consists of 169 items and 9k transactions.
    
    \subsection{Baseline Caching Algorithms}
        Since \spaarc{} is complementary to the underlying cache algorithm, we consider four native cache eviction policies in our evaluation: FIFO, LRU, LFU and Popularity (POP). Comparison is carried out {\em with and without} \spaarc{} integration with the cache. 
        Unless specified, we are not tuning \spaarc{} in the experiments in this section. The results of auto-tuning algorithm are presented separately.

    \subsection{Benefit of \spaarc{} integration}
        \begin{table}
            \caption{Percentage hit rate improvement of \spaarc{} over baseline}
            \label{tab:percenthr}
            \begin{center}
                \begin{tabular}{|l||*{4}{c|}}\hline
                    \diagbox[height=2\line]{Dataset}{Cache}& LRU & LFU & FIFO & POP\\
                    \hline\hline
                    DS30  & 31.15 & 24.55 & 40.6 & 32.01\\
                    \hline
                    DS45  & 20.33 & 7.64 & 26.18 & 25.05\\
                    \hline
                    DS60  & 24.44 & 12.28 & 27.56 & 32.4\\
                    \hline
                    DS75 & 12.87 & 3.27 & 22.44 & 9.65\\
                    \hline
                \end{tabular}
            \end{center}
        \end{table}
        
        This experiment evaluates \spaarc's effectiveness when integrated with cache eviction algorithms (baselines).  We varied both minimum support and minimum confidence from 30\% to 75\% while keeping other parameters constant (association factor at 1, and proximity threshold at 15). As shown in Figure \ref{fig:besthr} and Table \ref{tab:percenthr}, \spaarc{} significantly improves hit rates compared to baselines by 3.27\% to 40.6\% across all datasets.

        \spaarc{} demonstrates the highest relative improvement over the FIFO baseline, particularly with the DS30 dataset. This highlights \spaarc's ability to mitigate FIFO's limitation of potentially evicting frequently accessed items by proactively caching those likely to be accessed in the near future. Similar improvements are observed for LRU and POP. However, LFU exhibits lower hit rates due to its tendency to evict recently added, associated items.
        
        Furthermore, hit rates generally increase from DS30 to DS75, indicating that both baseline algorithms and \spaarc{} benefit from a greater number of relevant associations. Subsequent results provide a detailed analysis of each parameter's impact within specific configurations\footnote{For space reasons, we present results with a subset of the datasets and baseline algorithms. The trends hold for the omitted results.}.

    \subsection{Impact of minimum support}
\begin{figure}[h]
            \centering
            \includegraphics[scale=0.5]{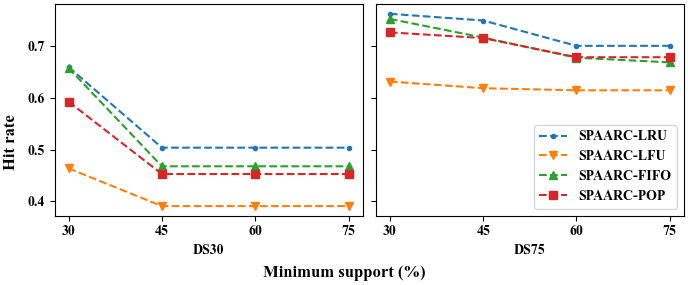}
            \caption{Varying minimum support at fixed minimum confidence of 45\%. }
            \label{fig:msup}
        \end{figure}

        This experiment investigates the impact of minimum support on \spaarc's performance. We varied the minimum support threshold from 30\% to 75\% while keeping other parameters constant (minimum confidence at 45\%, association factor at 1, and proximity threshold at 15). Figure \ref{fig:msup} shows the hit rate variation for DS30 and DS75, with similar trends observed for DS45 and DS60. This behavior is consistent across different minimum confidence values. As a critical parameter in association rule mining, higher minimum support thresholds result in fewer frequent itemsets and consequently, fewer association rules, potentially leading to lower hit rates.

    \subsection{Impact of minimum confidence}
\begin{figure}[h]
            \centering
            \includegraphics[scale=0.5]{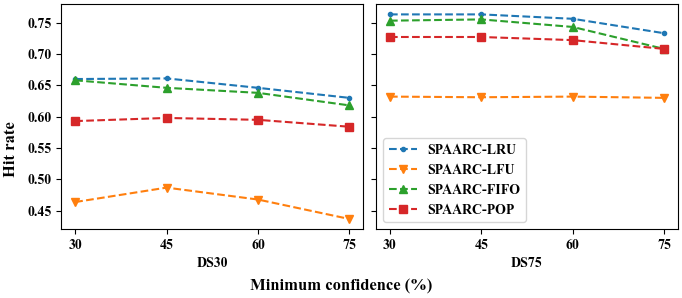}
            \caption{Varying minimum confidence at fixed minimum support of 30\%.}
            \label{fig:mconf}
        \end{figure}

        This experiment analyzes the influence of minimum confidence on \spaarc's performance. We varied the minimum support threshold from 30\% to 75\% while keeping other parameters constant (minimum confidence at 45\%, association factor at 1, and proximity threshold at 15). Figure \ref{fig:mconf} shows this effect on DS30 and DS75, with similar trends observed for DS45 and DS60 across different minimum support values. As a key parameter in ARM, increasing the minimum confidence threshold initially improves or maintains hit rates before causing a decline. This trend reflects the trade-off between rule quality and quantity. Higher thresholds reduce the number of rules, improving quality but potentially excluding relevant associations. 
        
    \subsection{Impact of Association factor threshold}
\begin{figure}[h]
            \centering
            \includegraphics[scale=0.4]{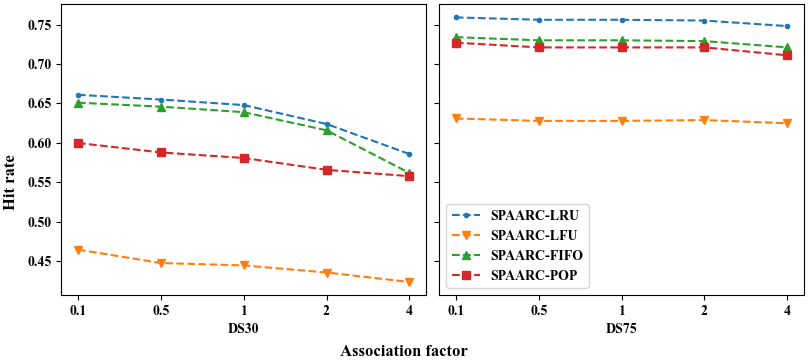}
            \caption{Varying association factor at a fixed minimum support of 30\% and minimum confidence of 45\%.}
            \label{fig:afactor}
        \end{figure}

        This experiment examines the impact of the association factor on \spaarc's cache hit rates. With minimum support and minimum confidence fixed at 30\% and 45\% respectively, the association factor was varied from 0.1 to 4. Figure \ref{fig:afactor} illustrates the results. As dataset support increases (i.e., associated objects are accessed more frequently), the influence of the association factor on identifying relevant objects diminishes.  For DS30, the highest hit rate is achieved with an association factor of 0.1, demonstrating a clear impact. However, for DS75, where frequent itemsets have higher support, the association factor has a less pronounced effect. This suggests that the association factor plays a more critical role in scenarios with lower object access frequencies.

    \subsection{Impact of Proximity with Association}
\begin{figure}[h]
                    \centering
                    \includegraphics[scale=0.4]{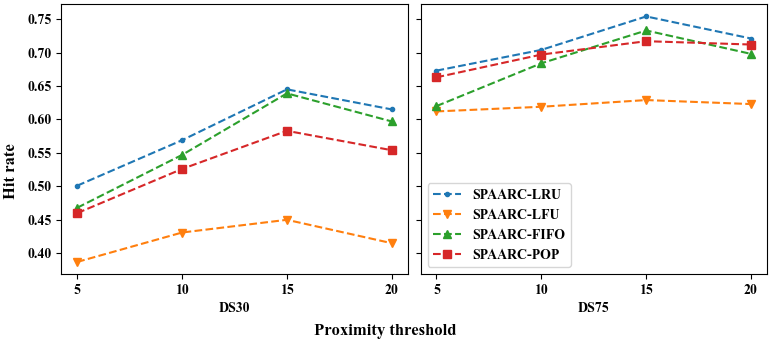}
                    \caption{Varying proximity threshold at a fixed minimum support of 30\% and minimum confidence of 45\%}
                    \label{fig:proximity}
                \end{figure}

            Figure \ref{fig:proximity} illustrates the influence of the proximity threshold on cache hit rates for \spaarc{}-integrated baselines. We varied the proximity threshold from 5 to 20 unit distance while fixing the minimum support at 30\%, minimum confidence at 45\%, and association factor threshold at 1. The proximity threshold has a significant impact on hit rates. A very low threshold might neglect relevant objects located slightly further away, potentially missing prefetching opportunities. Conversely, an excessively high threshold could lead to prefetching irrelevant objects that are not close enough for immediate user interaction, wasting cache resources. Therefore, selecting an appropriate proximity threshold is crucial for optimizing \spaarc{}'s effectiveness. Similar results are observed for DS45 and DS60.
    
\begin{figure}[h]
                    \centering
                    \includegraphics[scale=0.45]{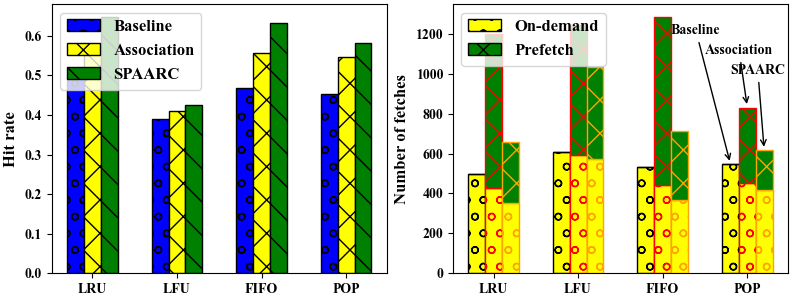}
                    \caption{Hit rate and fetch comparison across baseline, association-only prefetch and \spaarc{}. 
                    }
                    \label{fig:overhead_fetches}
                \end{figure}

        Figure \ref{fig:overhead_fetches} 
        shows the relative performance and overhead of using association only  and association+proximity (\spaarc) on top of the baseline caching algorithm, using the DS30 dataset with fixed parameters.
        The left graph shows that the hit rate improvement associated with association and \spaarc{} are 3- 20\% and 8- 36\% respectively. 
        The right graph measures the overhead of using \spaarc{} by comparing on-demand fetches and prefetches across the baseline, baseline with association-based prefetching, and baselines with \spaarc{}. 
        It shows that both association-based prefetching and \spaarc{} reduce on-demand fetches compared to the baseline by 2-17\% and 23-31\% respectively, as prefetching caches relevant objects. However, this introduces a prefetch overhead. Notably, association-based prefetching alone has a significantly higher overhead (0.51-2.12X) than \spaarc{} (0.12-0.70X) due to its lack of irrelevant object filtering. 
        Overall, the reduced on-demand fetches lead to improved hit rate and lower latency. While the prefetching occurs in the background without degrading user experience, it still leads to additional bandwidth consumption 
        which we plan to address through parameter tuning and encoding in future work.

    \subsection{Number of Users}
\begin{figure}[h]
            \centering
            \includegraphics[scale=0.5]{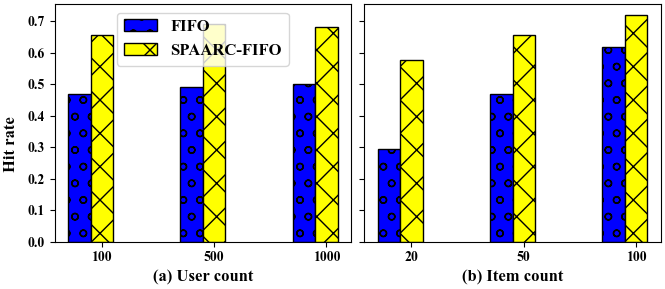}
            \caption{Varying number of users and items. 
            }
            \label{fig:vusers_vitems}
        \end{figure}

        This experiment evaluates the impact of user count on \spaarc's performance by varying the number of users from 100 to 1000 for FIFO cache policy. Figure \ref{fig:vusers_vitems}(a) shows the best hit rates achieved for DS30 across all minimum support and minimum confidence values.  \spaarc{}-FIFO improves hit rates by 5-40\% when compared to FIFO. Importantly, \spaarc{} maintains consistent hit rate improvements across the varying user counts in each dataset. This suggests that \spaarc's performance is robust to changes in user count.

    \subsection{Number of Unique Objects}
        This experiment examines the impact of virtual object count on \spaarc's effectiveness, a crucial factor varying significantly across application domains.  We varied the number of virtual objects from 20 to 100 for DS30 workload for 500 users on FIFO. Figure \ref{fig:vusers_vitems}(b) presents the best achievable hit rates for \spaarc{}-FIFO compared to FIFO.

        Results show a decreasing trend in hit rate improvement as object count increases within the same region. For instance, it is 96\% for 20 objects, 40\% for 50 objects, and 15\% for 100 objects.  Similar trends are observed in other datasets and cache policies. This can be attributed to increased object density, leading to more objects falling within the user's field of view (FoV).  Consequently, baseline caching policies are more likely to cache relevant objects by chance, reducing the relative advantage of \spaarc{}'s targeted prefetching strategy.

    \subsection{Minimum support tuning}
\begin{figure}[t]
            \centering
            \includegraphics[scale=0.4]{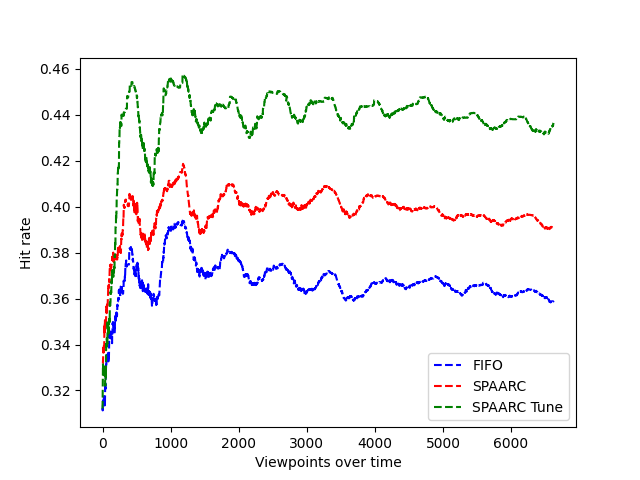}
            \caption{Minimum support tuning on synthetic workload. 
            }
            \label{fig:tuningsynth}
        \end{figure}

        In this experiment, we focus on tuning minimum support value with a fixed minimum confidence of 10\%, cache capacity of 10\%, hit rate degradation threshold of 5\%, initial transaction count for generating association rules of 100 and DS30 workload of 1000 users. Figure \ref{fig:tuningsynth} shows the hit rate associated with FIFO baseline, FIFO with \spaarc{} (no tuning) and FIFO with tunable \spaarc{}. It can be seen than dynamic tuning of \spaarc{} leads to higher hit rates. 
        \spaarc{}-Tune achieves 10\% and 22\% better hit rate compared to \spaarc{} and FIFO baseline. The hit rates are recorded every 10 views (1 viewpoint) and checked for degradation. If it is above a threshold, the association rule sets are generated. During the generation of association rules, incoming requests are served in parallel. Once the association rule set is selected, it is used for identify associated objects.
        
    \subsection{Real World experiments}
\begin{figure}[h]
            \centering
            \includegraphics[scale=0.4]{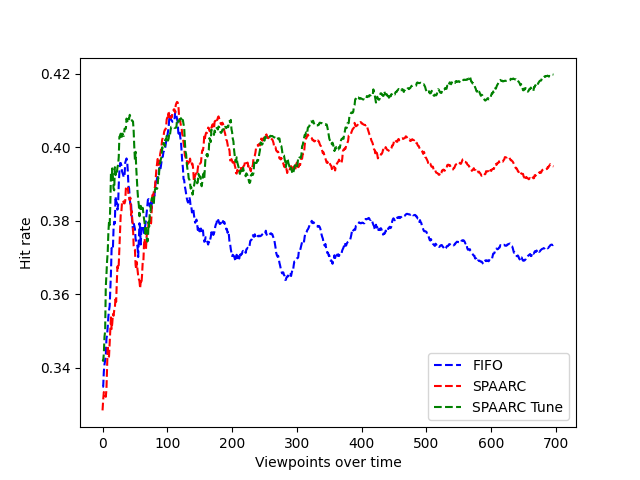}
            \caption{Minimum support tuning on real workload. \spaarc{} is able to perform 6\% and 11\% better compared to \spaarc{} and FIFO.}
            \label{fig:tuningreal}
        \end{figure}

        We use a real world dataset from SPMF \cite{bib:spmf} that consist of 169 items and 1000 transactions. We use Amazon Local zone for the edge server and AWS EC2 for the cloud server. The cache capacity is set to 10\%, hit rate degradation to 5\% and initial transaction count to 100 for rule generation. Figure \ref{fig:tuningreal} shows the hit rate performance. The hit rates are recorded every 100 views (1 viewpoint) and checked for degradation. It could be seen that \spaarc{} with dynamic tuning is able to perform better than \spaarc{} and FIFO as time passes by. Initially, it takes time to warm-up and then the hit rate starts increasing. 
        \spaarc{}-Tune achieves 6\% and 11\% better hit rate compared to \spaarc{} and FIFO baseline.
    \section{Related Work}
\label{sec:related}
    Mobile augmented reality (MAR) applications \cite{bib:arsurvey1} enhance user perception by integrating virtual objects into their view. These applications typically employ a distributed architecture across mobile devices, edge servers, and the cloud \cite{bib:edgearch, bib:arena, bib:iotarch}.  Efficiently partitioning compute- and data-intensive operations across these tiers is crucial. While MAR applications initially relied on local storage on mobile devices for rapid retrieval and rendering of virtual objects \cite{bib:localcache}, on-demand fetching from the cloud has become prevalent with increasing application complexity \cite{bib:cloudar1, bib:cloudar2, bib:cloudar3, bib:cloudar4}. However, this approach faces challenges in delivering immersive experiences due to increasing object sizes and high cloud latency.

    Edge caching has emerged as a promising solution for AR applications \cite{bib:cachemec, bib:artactile}, reducing latency and offering greater storage capacity compared to mobile devices.  Existing research explores various edge caching strategies. Cachier \cite{bib:cachier} employs a latency-minimizing model that balances load distribution between cloud and edge, considering network conditions and request locality. Agar \cite{bib:agar} uses dynamic programming to identify popular data chunks for caching. CEDC-O \cite{bib:onlinecoll} formulates edge data caching as an optimization problem considering caching cost, migration cost, and quality-of-service penalties.
    

    While CARS \cite{bib:carsar} and SEAR \cite{bib:sear} utilize Device-to-Device (D2D) communication for sharing cached virtual objects, they lack prefetching mechanisms. DreamStore \cite{bib:dreamstore} allows devices to store virtual objects and employs a publish-subscribe model for data updates, incorporating location-based prefetching but without considering object relevance. Precog \cite{bib:precog} and \cite{bib:arwmnedge} utilize Markov models for query prediction but do not account for inter-object relationships, a key aspect of \spaarc{}.
    
    While CARS \cite{bib:carsar} and SEAR \cite{bib:sear} leverage Device-to-Device (D2D) communication to share cached virtual objects with nearby users, they lack prefetching capabilities. DreamStore \cite{bib:dreamstore} allows user devices to store virtual objects and employs a publish-subscribe mechanism for data updates. It also incorporates a location-based prefetching mechanism, but fails to consider object relevance to the user. Precog \cite{bib:precog} and \cite{bib:arwmnedge} utilize a Markov model to predict user queries, but do not account for relationships between objects in a region, which is a key strength of \spaarc{}.

    Other research efforts focus on caching content for different media types at the edge. Works like \cite{bib:coopec, bib:tile,bib:cubist} explore caching tiles of 360$^\circ$ videos to enable processing reuse. Space \cite{bib:space} and Leap \cite{bib:leap} investigate prefetching video segments for users at the edge. EdgeBuffer \cite{bib:edgebuffer} leverages user mobility patterns across access points to prefetch data to anticipated locations.

    Association rule mining (ARM) \cite{bib:armorig, bib:arm} has established itself as a valuable technique for prefetching data in various domains, including e-commerce recommendation systems, fraud detection, and social network analysis. For instance, Mithril \cite{bib:mithril} leverages historical patterns of cache requests within cloud applications to derive item association rules using a variant of ARM called sporadic-ARM. Similarly, web prefetching, which involves caching web objects in anticipation of user requests, is a well-researched area \cite{bib:webcache}. However, the potential of ARM for prefetching in augmented reality (AR) applications remains largely unexplored.
    \section{Conclusion}
\label{sec:conclude}
This paper investigated the potential of prefetching in mobile augmented reality (MAR) applications to improve user experience by reducing latency.  To address the compute and data intensive nature of AR, we proposed \spaarc{}, a prefetching policy for edge AR caches that leverages object associations and location information to proactively fetch virtual objects from the cloud. \spaarc{} incorporates tunable parameters, including minimum support and minimum confidence, which can be adjusted to meet specific application needs. 
Through extensive evaluation using both synthetic and real-world workloads, we demonstrated that \spaarc{} significantly improves cache hit rates compared to standard caching algorithms, achieving gains ranging from 3\% to 40\% while reducing the need for on-demand data retrieval from the cloud. Further, we presented an adaptive tuning algorithm that automatically tunes \spaarc{} parameters to achieve optimal performance. Our findings demonstrate the potential of \spaarc{} to substantially enhance the user experience in MAR applications by ensuring the timely availability of virtual objects.  Future research could explore further refinement of the parameter tuning process and investigate methods to mitigate the communication overhead.



\bibliographystyle{IEEEtran}
\bibliography{IEEEabrv, refer}

\end{document}